\documentclass[a4paper, 12pt]{article}
\usepackage[utf8]{inputenc}
\usepackage[english]{babel}
\usepackage{amsmath, amssymb, amsfonts}
\usepackage{fullpage, indentfirst, cite}
\usepackage{epsfig, graphicx, graphics}
\usepackage{color}
\usepackage{hyperref}

\newcommand{\be}{\begin{equation}}
\newcommand{\ee}{\end{equation}}

\title{
{\Large Heavy decaying dark matter and large--scale anisotropy of high--energy cosmic rays}
\date{}
\author{O.~E.~Kalashev and M.~Yu.~Kuznetsov\footnote{mkuzn@inr.ac.ru}
\vspace{.2cm}\\
\footnotesize \it Institute for Nuclear Research of the Russian Academy of Sciences, \\
\footnotesize \it 60th October Anniversary Prospect 7a, 117312 Moscow, Russia}
}

\begin{document}

\begin{flushright}
INR-TH-2017-009
\end{flushright}

{\let\newpage\relax\maketitle}

\begin{abstract}	
We examine the role of the large--scale anisotropy of the high--energy
cosmic ray distribution in a search for the heavy decaying dark matter (DM) signal.
Using recent anisotropy measurements from the extensive air shower (EAS)
observatories we constrain the lifetime of the DM particles
with masses $10^{7}~\leq~M_X~\leq~10^{16}$~GeV. These constraints
appear to be weaker than that obtained with the high energy gamma--ray limits.
We also estimate the desired precision level for the anisotropy measurements
to discern the decaying DM signal marginally allowed by the gamma--ray limits and discuss the prospects
of the DM search with the modern EAS facilities.
\end{abstract}


\section{Introduction}
The hypothesis of dark matter (DM) consisting of heavy long--living particles
has attracted significant attention in the context of inflationary
cosmology~\cite{Kuzmin:1997jua, Berezinsky:1997hy}. There are
several scenarios of effective DM particles production on various stages
of early Universe evolution~\cite{Zeldovich:1971mw, Zeldovich:1977, Kuzmin:1997jua,
Berezinsky:1997hy, Kofman:1994rk, Khlebnikov:1996zt, Khlebnikov:1996wr,
Kuzmin:1998kk, Chung:1998rq, Chung:1998zb, Kuzmin:1998uv}.
Although, heavy DM was discussed in other
contexts as well~\cite{Khlopov:1987bh, Fargion:1995xs, Gondolo:1991rn}.
From the experimental point of view the most appropriate method to
search for such DM particles is to look for the secondary high--energy
cosmic--ray flux from the particle decay. Historically the first indication on super--heavy DM existence 
came from the observation of super--GZK cosmic rays in AGASA~\cite{Takeda:2002at}.
However later on the GZK cut--off existence was confirmed by the next generation cosmic
ray experiments~\cite{AbuZayyad:2012ru, Abraham:2008ru}. Several DM decay based
interpretations~\cite{Murase:2015gea, Bhattacharya:2014vwa, Esmaili:2013gha, Dev:2016qbd, Esmaili:2014rma, Rott:2014kfa}
have been proposed for the detection of PeV neutrinos in IceCube~\cite{Aartsen:2013jdh, Aartsen:2014gkd}.
While most of these interpretations are disfavored by the recent studies in which 
the respective gamma-ray signal~\cite{Kuznetsov:2016fjt, Cohen:2016uyg} is analyzed, some DM models
with suppressed photon production are still viable~\cite{Feldstein:2013kka, Dev:2016qbd, Hiroshima:2017hmy}.

Technically the heavy DM candidate $X$ has two main parameters: mass
$M_X$ and lifetime $\tau$. Absolutely stable $X$--particles
are not so interesting from the experimental point of
view since its annihilation cross--section is bounded by unitarity:
$\sigma_X^{\rm ann.} \sim 1/M_{X}^2$, which makes its indirect detection
impossible for the todays experiments~\cite{Gorbunov:2011zz}.
There are several sources of constraints for the heavy
DM parameters. The mass is subjected to cosmological
constraints~\cite{Kolb:1998ki, Kuzmin:1998kk, Kuzmin:1999zk,
Chung:1998zb, Chung:2004nh, Gorbunov:2012ij},
and the lifetime of the DM particles can be effectively constrained
with the observed fluxes of various high--energy particles or with the upper limit on these fluxes.
For example, in Ref.~\cite{Kalashev:2008dh} the constraints have been put using the shape of charged
cosmic--ray spectra. However, with the modern cosmic ray data this method is not as effective in constraining $\tau$
as gamma--ray~\cite{Murase:2012xs, Cohen:2016uyg, Aloisio:2015lva, Esmaili:2015xpa, Kalashev:2016cre}
and neutrino~\cite{Cohen:2016uyg, Kuznetsov:2016fjt} data.

Another observable sensitive to the heavy DM decay
is the cosmic--ray anisotropy. Apart from the DM searches,
the anisotropy is a powerful tool for the elucidation of the cosmic--ray origin and propagation.
In particular it is useful in study of the galactic magnetic field structure imprinted
in cosmic rays~\cite{Ahlers:2016njd, Ahlers:2013ima, Mertsch:2014cua, Giacinti:2011mz}
or in search for the extended cosmic--ray sources such as large scale
structure~\cite{Sigl:2003ay, Sigl:2004yk, Kalashev:2007ph, Koers:2008ba}.
While for TeV--PeV energies the existence of large--scale anisotropy has been confirmed by several
experiments~\cite{Aglietta:2009mu, Aartsen:2016ivj}, for higher energies
the deviations from isotropy are either not observed or have low
significance~\cite{Antoni:2003jm, Chiavassa:2015jbg, Aab:2014ila, Ivanov:2014soa,
Aab:2016ban, AlSamarai:2015hwx}.
In the present work we use the upper--limits on the cosmic ray flux anisotropy mentioned above to 
obtain the conservative constrains on the lifetime of the heavy DM with masses
$10^{7}\leq~M_X\leq~10^{16}$~GeV. We use the parameters of DM allowed by the gamma--ray and neutrino
limits to reveal the possible DM contribution to the anisotropy observables.
This study complements our previous works~\cite{Kalashev:2016cre, Kuznetsov:2016fjt},
where constraints on the heavy decaying DM lifetime were obtained using 
the high--energy gamma--ray and neutrino flux upper limits.

\section{Cosmic ray flux from the dark matter decay}
\label{flux}
We consider DM consisting of scalar particles $X$
decaying through the primary channel $X \rightarrow q \bar{q}.$
This implies hadronisation and subsequent decay of unstable hadrons.
The final products of the decay cascade are photons, protons, neutrino
etc. In this study we are interested in the decay products that can contribute
to the  cosmic--ray flux anisotropy observable at Earth --- that is
photons and protons. We follow the method of Refs.~\cite{Aloisio:2003xj, Sarkar:2001se}
in calculation of the decay spectra. The details were reviewed in our previous
works~\cite{Kalashev:2016cre, Kuznetsov:2016fjt} along with the justification
of the approximations used. Here we just list
the key points. We assume that all quark flavors are coupled to $X$ similarly.
The photon production is assumed to be saturated by the pion decays while
the $\sim 10\%$ contribution from kaons and other mesons is neglected.
We also neglect the electro--weak corrections to the decay spectrum and
assume that final state particles momenta are distributed isotropically on average.

\begin{figure}
   \includegraphics[width=13.50cm]{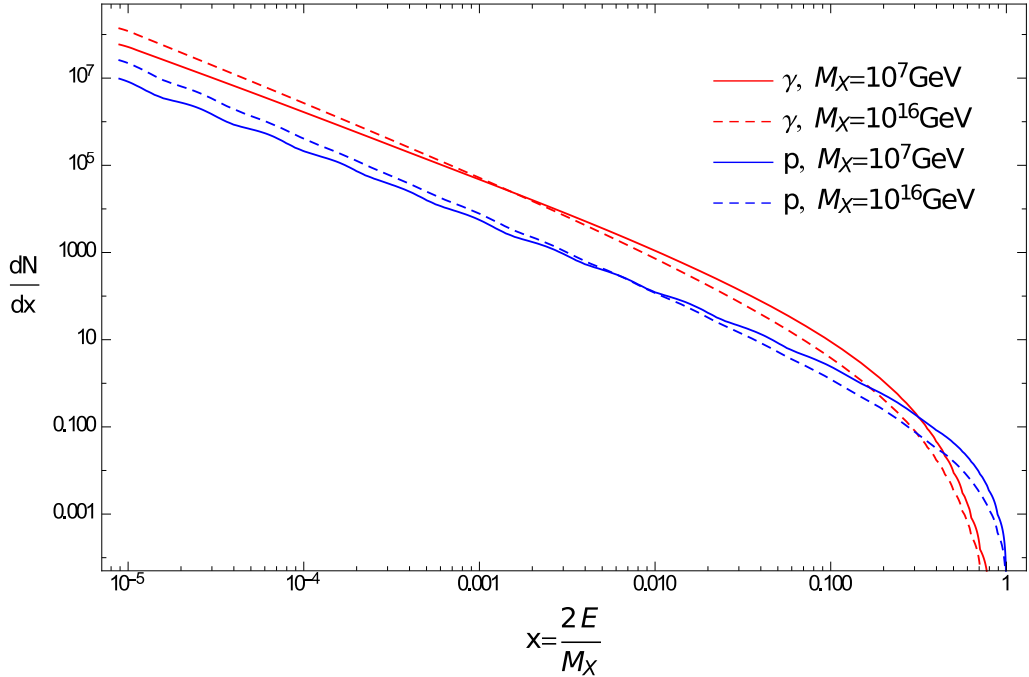}
   \caption{The spectra of photons and protons from $X$ particle
    decay for two different values of $M_X$ as a function of dimensionless
    variable $x=\frac{2 E}{M_X}$.}
   \label{decay_spectra}
\end{figure}

The spectra of photons and protons from the decay of particle
of mass $M_X$ can be obtained by the DGLAP
evolution~\cite{Gribov:1972ri,Lipatov:1974qm, Dokshitzer:1977sg,Altarelli:1977zs} of the low energy scale
fragmentation functions. For this calculation we use the code provided by the
authors of Ref.~\cite{Aloisio:2003xj} that solves DGLAP equations numerically in
the leading order of $\alpha(s)$. It assumes that all quark flavours are coupled to
gluon similarly and implies the mixing of gluon fragmentation function with the quark
singlet fragmentation function. The initial fragmentation functions parametrized on
scale of $1$~GeV are taken from Ref.~\cite{Hirai:2007cx} and extrapolated to
the interval $10^{-5}~\le~x~\le~1$. The examples of photon and proton spectra
from the decay of $X$ particles with various masses are shown in Fig.~\ref{decay_spectra}.

The large--scale anisotropy model predictions  should be calculated for the total flux of the
high--energy cosmic rays. It is known that the flux is dominated by the
isotropic contribution of charged particles which we denote as $J_{\rm exp}(E)$, while
the possible decay of the DM gives only a small anisotropic admixture denoted
by $J_{\rm DM}(\delta, \alpha, E)$, where $\{\delta, \alpha\}$ is equatorial coordinates.
The ``experimental'' flux is taken from Telescope Array~\cite{Ivanov:2015pqx}
and Tibet~\cite{Amenomori:2008aa} spectrum measurements. We choose these two experiments,
since their fluxes are close to the average of all the experimental spectra reported in Ref.~\cite{Agashe:2014kda}.
The uncertainty in anisotropy predictions due to discrepancy between spectra measured in different experiments
is estimated in the next section. We parametrize the ``experimental'' flux in the following way:
\be
\label{Jexp}
J_{\rm exp}(E)= J_1
\begin{cases}
	E_\text{-1}^{\gamma_{-1}-\gamma_{0}} E_{0}^{\gamma_{0}-\gamma_{1}} E^{-\gamma_{-1}} \quad E < E_{-1}\\	
	E_{0}^{\gamma_{0}-\gamma_1} E^{-\gamma_0} \quad E_\text{-1} \leq E < E_{0}\\
	E^{-\gamma_1} \quad E_\text{0} \leq E < E_\text{1}\\
	E_\text{1}^{\gamma_2-\gamma_1} E^{-\gamma_2} \quad E_\text{1} \leq E < E_{2}\\
	E_\text{1}^{\gamma_2-\gamma_1} E_{2}^{\gamma_3-\gamma_2} E^{-\gamma_3} \quad E > E_{2}\\
	\end{cases}\,,
\ee
where $E_{-1} = 4.0 \cdot 10^{15}$~eV, $E_{0} = 1.0 \cdot 10^{17}$~eV, $E_{1} = 5.2 \cdot 10^{18}$~eV
and $E_{2} = 6.3 \cdot 10^{19}$~eV corresponds to the energies of ``knee'', ``second knee'', ``ankle'' and
GZK cut-off respectively and the normalization factor $J_1$ is taken at $10^{18}$~eV. The values of spectral indexes
are: $\gamma_{-1}=2.72$, $\gamma_{0}=3.12$, $\gamma_{1}=3.23$,
$\gamma_{2}=2.66$, $\gamma_{3}=4.65$. 

In turn, the DM part of the flux consists of proton and photon
contributions~\footnote{We neglect the comparable neutrino flux because the
sensitivity of EAS experiments to neutrino are at least two orders of magnitude
smaller than that to photons and protons.
}.
\be
\label{full_flux}
J_{\rm DM}(\delta, \alpha, E) = J_p^{\rm G}(\delta, \alpha, E) + J_\gamma^{\rm G}(\delta, \alpha, E) +
J_p^{\rm EG}(E) + J_\gamma^{\rm EG}(E)
\ee
where G stands for ``Galactic'' and denotes contribution from the DM decay in Milky Way
while EG stands for ``Extra--Galactic'' and denotes the contribution from the rest of Universe.
Since here we discuss DM decay and not interesting in annihilation the anisotropic 
patterns related to the matter clustering in the extragalactic contribution are
washed out~\cite{Cirelli:2010xx} and this contribution can be considered as isotropic.
For the purposes of anisotropy study it is convenient to consider the extragalactic DM contributions
as a part of $J_{\rm exp}(E)$.  Indeed, at low energies where $J_{\rm exp}$ dominates over $J_{\rm DM}$
the predicted anisotropy is small and the accounting of $J_{\rm DM}^{\rm EG}$ provides only few percent correction to it.
Above the GZK threshold energy the $J_{\rm DM}$ starts to dominate, but extragalactic
photon and proton DM fluxes are suppressed by attenuation effect in the same way as
$J_{\rm exp}$ and only galactic contributions are relevant.
These assumptions were justified by direct calculations of anisotropy with the actual parameters
of the DM. Therefore, we take the total flux in the following form
\be
J_{\rm tot}(\delta,\alpha,E) = J_{\rm exp}(E) + J_p^{\rm G}(\delta, \alpha, E) + J_\gamma^{\rm G}(\delta, \alpha, E)
\ee

\begin{figure}
   \includegraphics[width=13.50cm]{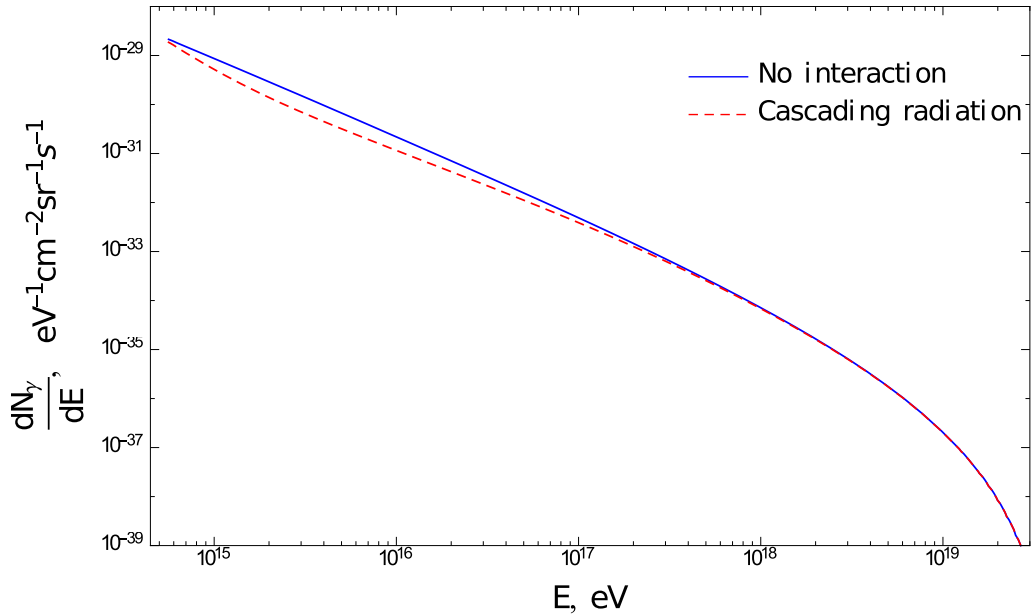}
   \caption{Predictions for the observable photon spectrum made with (dashed line)
   and without (solid line) contribution of the cascade on cosmic photon background for the DM decay model
   with mass $M_X=10^{11}$~GeV and lifetime $\tau = 10^{20}$~yr.}
   \label{noint_vs_cascade}
\end{figure}

For the galactic flux calculation we use the Navarro-Frenk-White DM
distribution~\cite{Navarro:1995iw, Navarro:1996gj} with the parametrization for Milky Way 
from  Ref.~\cite{Cirelli:2010xx}\footnote{For comparison we also calculate the resulting
anisotropy using Burkert DM profile~\cite{Burkert:1995yz}.}. For galactic gamma--ray flux we take into account
only prompt photon spectra of DM decay and neglect the smaller
amount of photons from inverse Compton scattering (ICS) of prompt $e^\pm$ on the interstellar background photons.
This assumption was discussed in our previous paper~\cite{Kalashev:2016cre}.
However, we allow for the modification of photon spectra due to interactions with CMB photons.
This correction becomes important for the $E_\gamma \lesssim 10^{19}$~eV
i.e. for $M_X \lesssim 10^{14}$~GeV. We use numerical code~\cite{Kalashev:2014xna} which simulates
development of electron-photon cascades on CMB driven by the chain of $e^\pm$ pair production and
inverse Compton scattering. Although the code allows to calculate
the flux of the cascade photons it doesn't take into account deflections of
$e^\pm$ by the halo magnetic field. Since electrons in the code propagate
rectilinearly they produce less cascade photons. Therefore the calculated flux of photons
should be considered as conservative lower bound. 
The code also includes attenuation of photons on extragalactic background light (EBL),
though the effect of EBL is negligible on distances which we consider.
The corrections related to the production of electro--magnetic cascades
are important for the energies of photon lower than $\sim 10^{19}$~eV
--- above $10^{19}$~eV the correction is less then $1\%$. The comparison of cascading
spectrum of photons with the injected spectrum both calculated for the decay of DM
in Milky Way and propagated to Earth is presented in Fig.~\ref{noint_vs_cascade}.
In turn, the galactic proton contribution is affected by the galactic magnetic field, which
deflects the protons and therefore washes out the anisotropy pattern.

The Milky Way magnetic field can be decomposed to regular (large scale) and random
components~\cite{Haverkorn:2014jka}. The  large-scale magnetic field obtained from Faraday rotation
of pulsars and extragalactic sources is typically around 1.5-2$\mu$G. The total magnetic field in
the Solar neighborhood is about 6~$\mu G$, which suggests presence of comparable random component.
Towards the Galactic center, the magnetic field strength increases, reaching values 7.6-11.2~$\mu$G
at a Galactocentric radius of 4~kpc. The magnetic field strength in the gaseous halo, or thick disk,
is comparable to that in the disk, with an uncertainty of a factor 2-3. 
The Larmor radius of a particle with energy $E$ and electric charge $qe$ in a regular magnetic field is
\be
R_g=\frac{E}{qeB_\perp}
\simeq 1.1\times\;\frac{1}{q}\left({E\over 10^{18}{\,{\rm eV}}}\right)
\left({B_\perp\over{\rm \mu G}}\right)^{-1}{\,{\rm kpc}}\,,
\ee
where $B_\perp$ is the field component perpendicular to the particle's motion. The critical energy
$E_c$ for protons in the Milky Way magnetic field i.e. the energy where the Larmor radius equals to
the coherence length of the turbulent component, is estimated as $E_c\simeq 0.3$~EeV. The flux of
protons with energies $E<E_c$ is completely isotropic due to randomization of their momenta directions
by the turbulent magnetic field component, while protons with higher energies spiral around the regular
component of the field. Below we  conservatively assume that only protons with energies above $10^{19}$~eV
contribute to the flux anisotropy. We assume rectilinear propagation of these protons and neglect possible
contribution of lower energy protons to the anisotropy. We justify this approximation in the next section
by the comparison of proton and photon contributions to the anisotropy.

\section{Anisotropy analysis and results}
\label{results}
Several observables are used in the literature to study the large--scale angular distribution
of cosmic--rays. The most commonly used method is based on the cosmic ray flux harmonic analysis.
Unfortunately, none of the currently running EAS experiments is observing the full sky.
However the experiments with the full duty cycle cover some band on celestial sphere
due to the Earth rotation. In this situation the appropriate solution is the
Fourier analysis in right ascension, where the flux assumed to be the average in declination
over the particular experiments field of view. The flux can be presented in the form:
\be
J(\alpha, E) = A_0(E) + \sum_n \left[ A_n(E) \sin(n \alpha) + B_n(E) \cos(n \alpha) \right] \; ,
\ee
where
\begin{gather}
\label{fourier_coeffs} 
A_0(E) = \frac{1}{2\pi} \int\limits_{-\pi}^{\pi} J(\alpha, E) d\alpha \;,\\
A_n(E) = \frac{1}{\pi} \int\limits_{-\pi}^{\pi} J(\alpha, E) \cos(n\alpha) d\alpha \;,\\
B_n(E) = \frac{1}{\pi} \int\limits_{-\pi}^{\pi} J(\alpha, E) \sin(n\alpha) d\alpha \;.
\end{gather}
Below we use the  normalized coefficients, $a_{n} \equiv 2 A_{n}/A_{0}$,
$b_{n} \equiv 2 B_{n}/A_{0}$. 
The observable commonly reported by the experiments
is the normalized amplitude of the first harmonic:
\be
\label{dipole_amplitude}
r_1 = \sqrt{a_1^2 + b_1^2}
\ee
To obtain the theoretical prediction for this quantity one needs to take into
account the effective exposure of the particular experiment which is given by~\cite{Sommers:2000us, Aab:2014ila}:
\be
\omega(a, \delta,\theta_\text{max}) \sim (\cos a\,\cos\delta\,\sin\alpha_m+\alpha_m\sin a\,\sin\delta),
\ee
where $a$ is geographical latitude of the observatory, $\theta_\text{max}$ is the maximal zenith
angle accessible for fully efficient observation in the experiment and $\alpha_m$ is given by
\be
\alpha_m=\begin{cases}
0 & ;\xi>1,\\
\pi & ;\xi<-1,\\
\arccos\xi & ; -1 < \xi < 1\,;
\end{cases}
\ee
\be
\xi = \frac{(\cos\theta_\text{max}-\sin a\,\sin\delta)}{\cos a\,\cos\delta}.
\ee
After inclusion of the exposure into the analysis we have for $a_1$:
\be
a_1(E) = \frac{2 \int\limits_{\Omega} J_{\rm DM}(\delta, \alpha, E) \omega(a, \delta,\theta_\text{max})
\cos(\alpha) d\Omega}{J_{\rm exp}(E) \int\limits_{\Omega} \omega(a, \delta,\theta_\text{max}) d\Omega + 
\int\limits_{\Omega} J_{\rm DM}(\delta, \alpha, E) \omega(a, \delta,\theta_\text{max}) d\Omega} \;
\label{a1}
\ee
and for $b_1$:
\be
b_1(E) = \frac{2 \int\limits_{\Omega} J_{\rm DM}(\delta, \alpha, E) \omega(a, \delta,\theta_\text{max})
\sin(\alpha) d\Omega}{J_{\rm exp}(E) \int\limits_{\Omega} \omega(a, \delta,\theta_\text{max}) d\Omega + 
\int\limits_{\Omega} J_{\rm DM}(\delta, \alpha, E) \omega(a, \delta,\theta_\text{max}) d\Omega} \;.
\label{b1}
\ee
From expressions~(\ref{a1})-(\ref{b1}) it is easy to estimate the uncertainty of theoretically
predicted values of $a_1$ and $b_1$ due to variation of the experimental flux $J_{\rm exp}(E)$.
Generally, at lower energies $J_{\rm exp}(E)$ dominates over $J_{\rm DM}$, therefore the flux
error linearly maps to the uncertainty of $a_1(E)$ and $b_1(E)$, while at the higher energies $J_{\rm DM}(E)$
starts to supersede the ''experimental'' flux and the impact of its uncertainty
on the resulting anisotropy decreases. However, at $E \simeq 10^{20}$~eV, where
$J_{\rm DM}$ is still subdominant the experimental flux uncertainty is almost
two orders of magnitude.~\footnote{We estimate the uncertainty by comparison of the flux
measurements in Telescope Array and Pierre Auger experiments.} Therefore, the predictions
for anisotropy above $E \simeq 10^{20}$~eV should be interpreted with these reservations.

The way to improve the sensitivity of the experiments to large--scale anisotropy is to
consider summarized data of two experiments with fields of view jointly covering
the whole celestial sphere. In that case one can expand the flux in spherical harmonics.
This method has been adopted to study the anisotropy of the ultra-high energy cosmic rays
by Pierre Auger and Telescope Array experiments in Ref.~\cite{Aab:2014ila}~\footnote{In
recent studies of Pierre Auger~\cite{Aab:2016ban} and IceCube\cite{Aartsen:2016ivj}
the advanced analysis techniques was used to extract the full--sky angular
power spectrum from the partial sky data of these experiments. However, the restrictions
on the anisotropy patterns that could be extracted by these techniques makes questionable
its applicability to the DM decay anisotropy search.}.
The expansion of the flux into spherical harmonics has the form:
\be
J(\delta, \alpha, E) = \sum\limits_{l\geq 0}\sum\limits_{m=-l}^{l} a_{lm}(E) {\rm Y}_{lm}(\delta, \alpha)
\ee 
with the coefficients defined as
\be
\label{laplace_coeffs} 
a_{lm}(E) = \int\limits_{\Omega} J(\delta, \alpha, E) {\rm Y}_{lm}(\delta, \alpha) d\Omega \;,
\ee
where integration goes over the whole celestial sphere and the exposure effects assumed
to be eliminated from the resulting experimental quantities.
The quantity analogous to $r_n$ is the angular power spectrum
which is defined as
\be
\label{power_spectrum}
C_l=\frac{1}{2l+1}\sum_m |a_{lm}|^2 \; .
\ee
Since here we are interested anisotropy only, below we redefine $a_{lm} \rightarrow \sqrt{4\pi} a_{lm}/a_{00}$,
i.e. normalize coefficient to the monopole one.

The alternative approach for the DM signal search in the full--sky cosmic--ray map
would be the fitting of the map with the signal plus background model, using the
profile likelihood method. This technique was effectively employed by Fermi-LAT collaboration
for the search of MeV--GeV gamma-ray flux of DM decay/annihilation origin~\cite{Ackermann:2012rg}.
This method would be in general more sensitive to the DM signal than the harmonic analysis.
However, as the appropriate analysis tools are yet undeveloped, we leave this issue
for the future studies.

For any mass $M_X$ the lifetime $\tau$ can be constrained using the upper--limits on the amplitudes
(\ref{dipole_amplitude}, \ref{power_spectrum}) or on the coefficients (\ref{laplace_coeffs}).
We use the data from EAS-TOP~\cite{Aglietta:2009mu}, IceCube~\cite{Aartsen:2016ivj}, KASCADE~\cite{Antoni:2003jm},
KASCADE-Grande~\cite{Chiavassa:2015jbg}, Yakutsk~\cite{Ivanov:2014soa} and Pierre Auger~\cite{AlSamarai:2015hwx}
experiments. All data is interpreted in terms of $r_1$ amplitude $C.L.=95 \%$ upper-limits,
except the KASCADE-Grande data which is presented as $C.L.=99 \%$ upper-limit.
We also employ the result of joint Telescope Array and Pierre Auger full--sky anisotropy
study~\cite{Aab:2014ila} presented in the form of separate upper--limits on $a_{lm}$ coefficients.
We conservatively assume that all the anisotropy is given by the DM decay.
To obtain the constraints we vary the DM lifetime $\tau$ until the amplitude
(\ref{dipole_amplitude}) touches one of the experimental constraints from below.
The procedure for $a_{lm}$ limits is similar but we use all the coefficients $a_{lm}$
for given energy bin. To obtain 95 \% CL limit taking into account the
respective statistical penalty we use $(0.95)^{\frac{1}{2l+1}}$ CL limits derived from
all $(2l+1)$ coefficients $a_{lm}$.
Since the values of $a_{lm}$ lie in the sign-changing band in each case we
chose the edge of the band which has the sign similar to the respective
theoretically predicted coefficient as a limit value.

\begin{figure}
   \includegraphics[width=13.50cm]{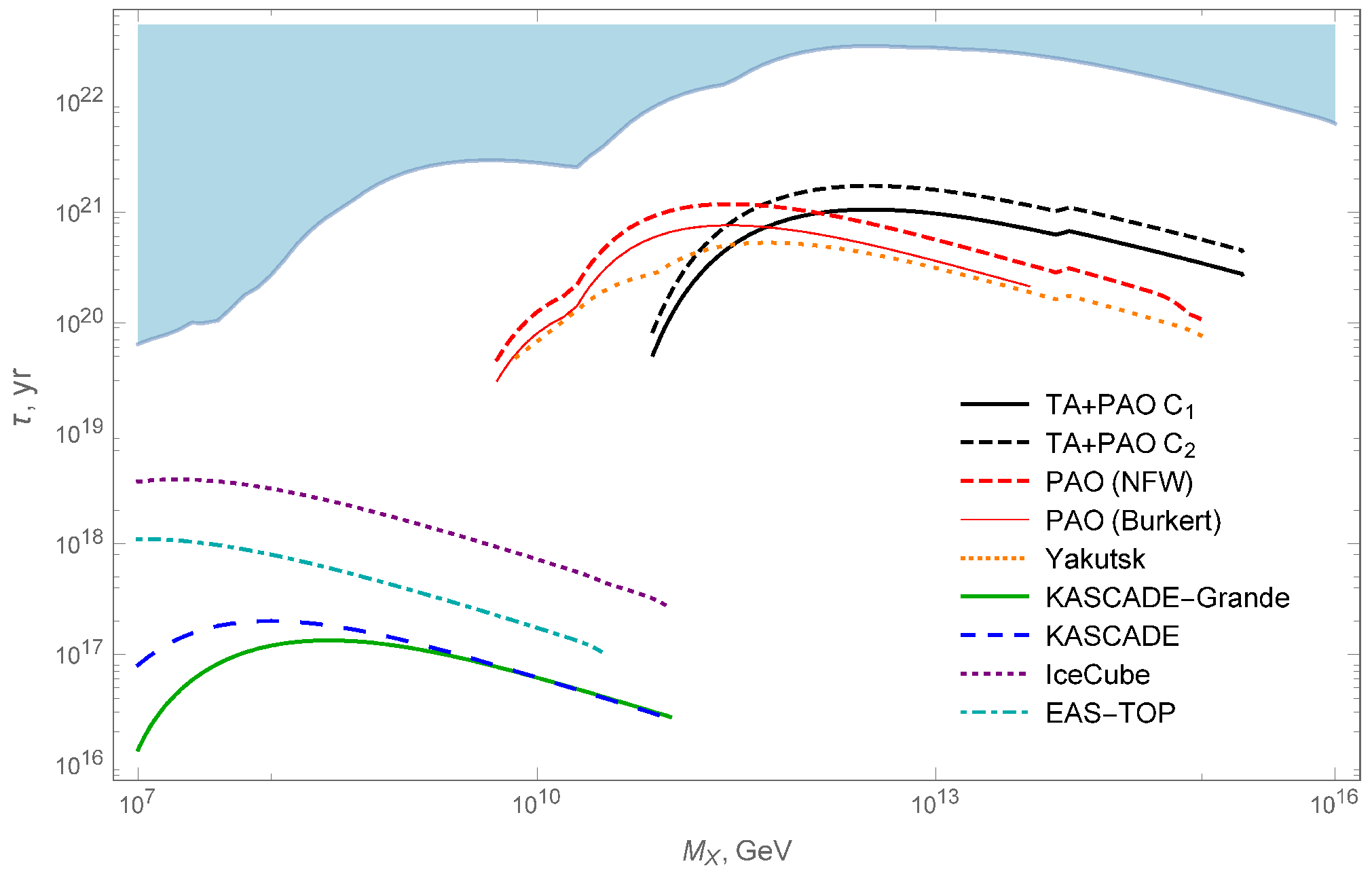}
   \caption{$95\%$ C.L. exclusion plot for mass $M_X$ and lifetime $\tau$ of DM particles.
   The constraints are obtained assuming NFW DM profile with the data of Telescope Array and
   Pierre Auger full--sky analysis~\cite{Aab:2014ila} first harmonic (solid black) and second
   harmonic (dashed black); data of Pierre Auger partial--sky analysis~\cite{AlSamarai:2015hwx};
   data of Yakutsk~\cite{Ivanov:2014soa} (dashed orange), IceCube~\cite{Aartsen:2016ivj} (dashed purple),
   EAS-TOP~\cite{Aglietta:2009mu} (dot--dashed cyan), KASCADE~\cite{Antoni:2003jm} (dashed blue)
   and  KASCADE-Grande~\cite{Chiavassa:2015jbg} (solid green) partial--sky analysis
   (for KASCADE-Grande C.L. is $99\%$). White area is excluded by the photon
   and neutrino constraints obtained in~\cite{Kalashev:2016cre, Kuznetsov:2016fjt}.
   Also we show for comparison the constraints obtained assuming Burkert DM profile (solid red)
   using the data of Pierre Auger partial--sky analysis~\cite{AlSamarai:2015hwx}}
   \label{exclusion_plot}
\end{figure}

The results are shown in Fig.~\ref{exclusion_plot}. As one can see
all the anisotropy constraints lie in the parameter area which is already excluded by the
high--energy gamma--ray and neutrino based limits. Since the anisotropy and gamma--ray
limits are set by the same experiments the above result indicates that EAS experiments are more sensitive
to the DM decay photons than to the respective anisotropy.
This fact has some interesting consequences that we discuss below.

Another important feature is that the anisotropy constraints are stronger for
the higher energies. This fact can be understood if we notice that the background isotropic
cosmic--ray flux grows faster with the decrease of energy than the precision
of anisotropy measurements. As it was anticipated the full--sky constraints surpass that of the
limited sky coverage. The surprising fact is that second harmonic of full--sky analysis bounds
$\tau$ stronger than the first one, although the theoretically predicted
amplitude of the former is generally smaller (see below). This is the effect of the
incidentally small value of the upper limit for one of the coefficients which is not compensated
by the statistical penalty. The constrains by the third harmonic which are not shown in the
figure are weaker than those by the first and second harmonic. The small bump around
$M_X = 10^{14}$~GeV on each curve is due to the accounting of galactic proton flux
along with the photons, its impact on result is expectedly small.

\begin{figure}
   \includegraphics[width=13.50cm]{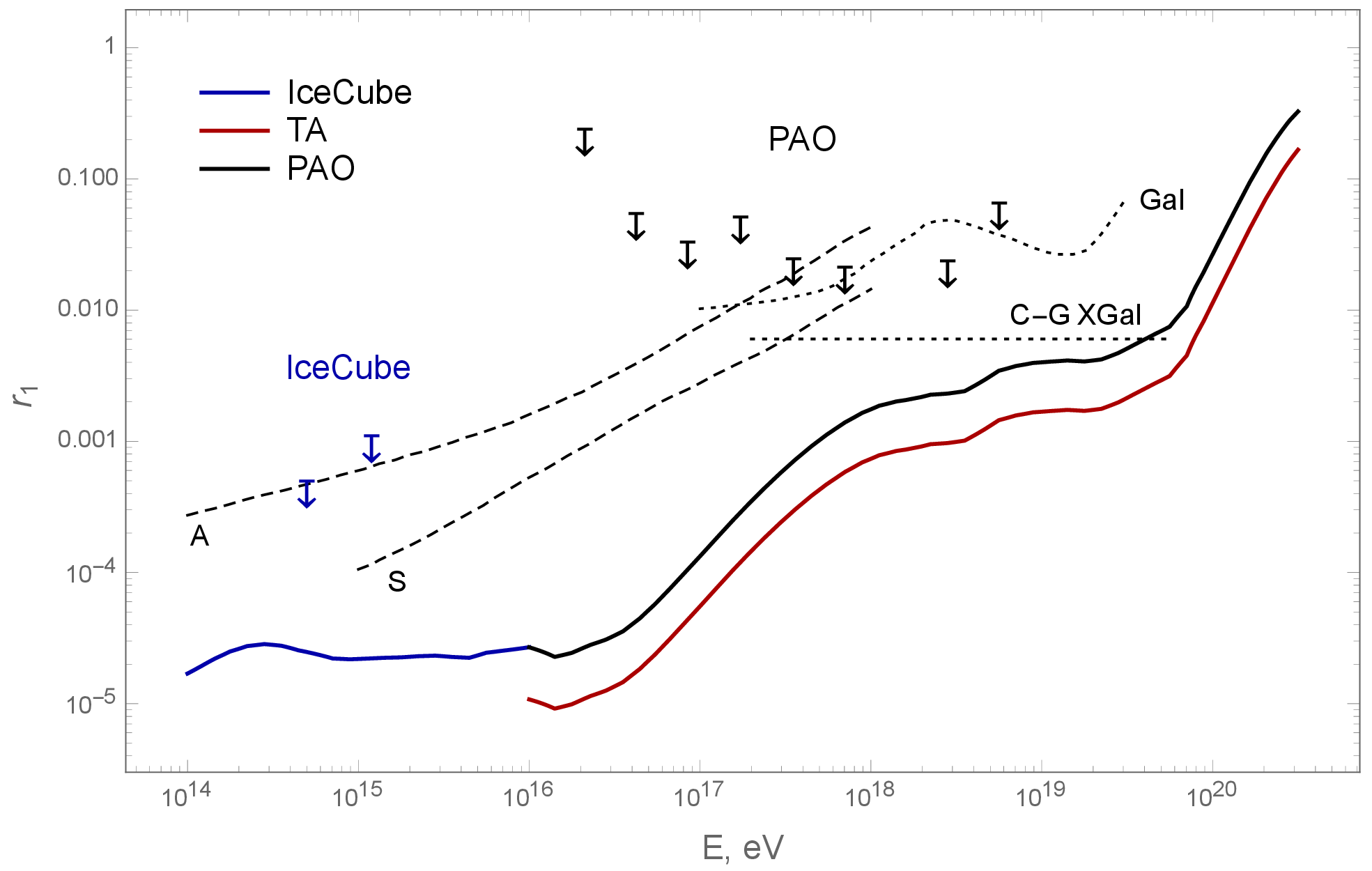}
   \caption{The expected cosmic ray flux anisotropy produced by the decay of DM with the
   lifetime marginally allowed by the gamma--ray constraints. The amplitude $r_1$ of the
   first harmonic of right ascension analysis is shown for IceCube, Telescope Array and
   Pierre Auger experiments. For each energy $E$ we calculate the value $r_1(E)$ in $0.1$
   decade wide energy interval and maximize it over all masses $M_X$ that can generate a flux at this energy.
   Recent limits from these experiments~\cite{Aartsen:2016ivj,AlSamarai:2015hwx}
   are shown for comparison. Also we show the predictions for the alternative scenarios of anisotropy origin:
   the predictions from two galactic magnetic field models with different symmetries (A and S)~\cite{Candia:2003dk},
   the predictions for a purely galactic origin of cosmic rays (Gal)~\cite{Calvez:2010uh} and
   the expectations from the Compton-Getting effect for an extragalactic cosmic ray flux (C-G EG)~\cite{Kachelriess:2006aq}.}
  \label{proposal_r1}
\end{figure}

\begin{figure}
   \includegraphics[width=13.50cm]{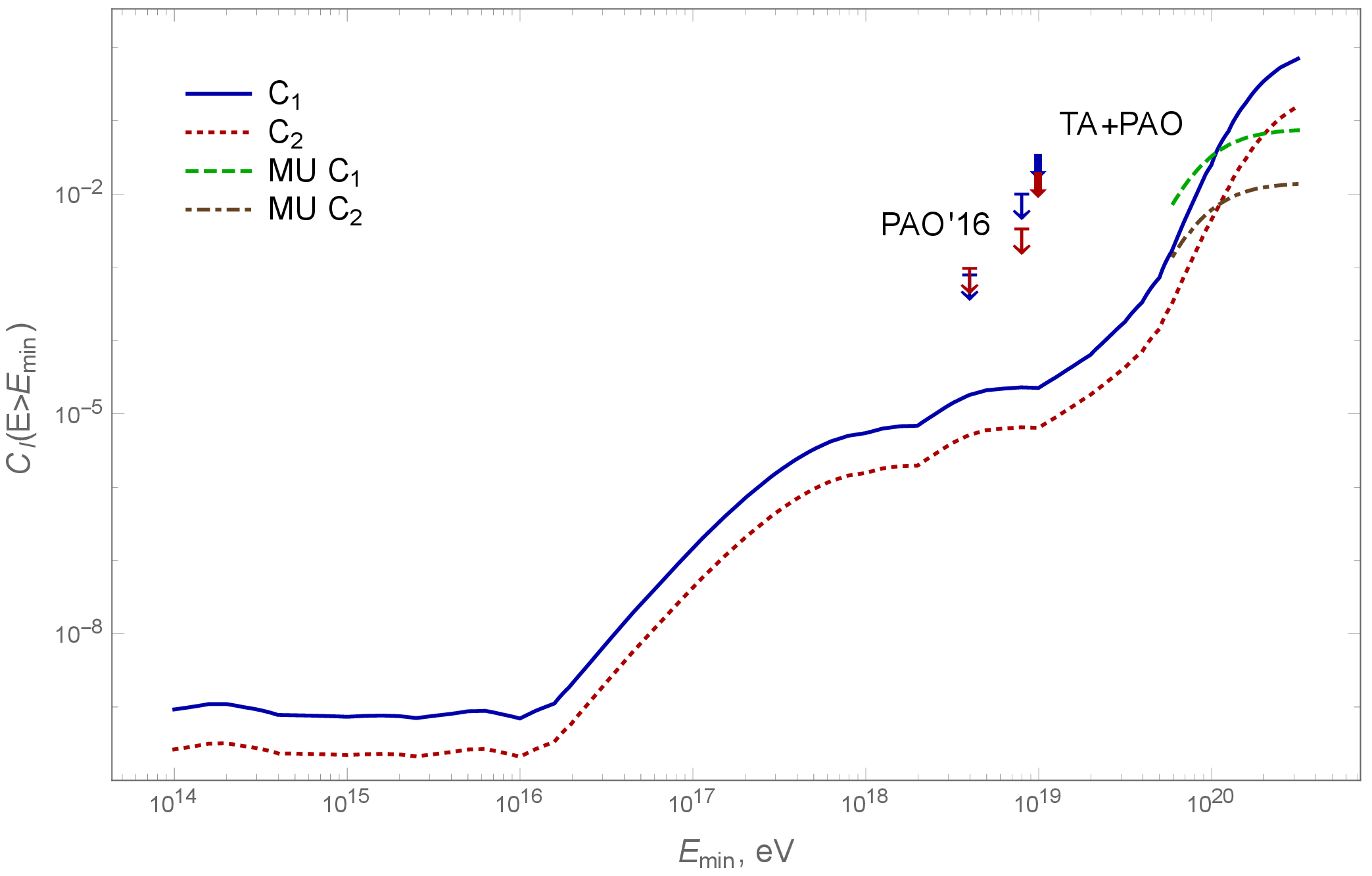}
   \caption{Full--sky anisotropy produced by the decay of DM with the lifetime marginally allowed
   by the gamma--ray constraints in terms of first and second angular power spectrum coefficients. For each energy
   $E_{\rm min}$ we show the integral values $C_1(E > E_{\rm min})$ (solid blue) and
   $C_2(E > E_{\rm min})$ (dotted red) maximal over all masses $M_X$ that can
   generate a flux at this energy. Similar predictions are shown for $C_1$ (dashed green)
   and $C_2$ (dot--dashed brown) obtained in Ref.~\cite{Marzola:2016hyt}.
   Experimental limits from TA and Auger joint analysis~\cite{Aab:2014ila}
   and from recent Auger study~\cite{Aab:2016ban} are shown for comparison.}
   \label{proposal_fullsky}
\end{figure}

For further experimental analysis development it is worth to discuss the maximal expected
large--scale anisotropy from DM decay allowed by the most recent gamma--ray
constraints~\cite{Kalashev:2016cre}. We obtain it for the particular experiments
and for the full--sky measurements by fixing the value of $\tau$ which is marginally
allowed by the gamma--ray limits. The results for the range of masses $M_X$ are shown in
Figs.~\ref{proposal_r1}--\ref{proposal_fullsky} together with the recent limits. For the
individual experiments we calculate the desired anisotropy sensitivity in terms of $r_1$.
Variation of $r_1$ with the choice of an experiment reflects the fact that
the anisotropy observed in particular experiment depends on its field of view.
We use Pierre Auger and Telescope Array for medium and high energies and IceCube for lower energies.
Increase in sensitivity and expanding of the energy range is expected in Telescope Array
with the planned deployment of the low energy extension TALE~\cite{Sagawa:2016ysc},
which will allow to collect events with the energies down to $10^{16}$~eV. The high area
extension TA$\times$4~\cite{Sagawa:2015yrf, Sagawa:2016ysc} should give significant increase
of the statistics at higher energies. One should note that the Southern hemisphere based
experiments --- Pierre Auger and IceCube have an advantage in galactic anisotropy study
over the Northern hemisphere based Telescope Array because of more convenient position
relative to the Galactic Center. We see that significant increase in experimental sensitivity
to the large--scale anisotropy is need to detect the maximal signal expected from the DM decay. The
IceCube sensitivity should be increased by roughly two orders of magnitude,
while for Auger the respective values are from $\sim 10^4$ to $\sim 20$ times at 1~EeV energy.

For the full--sky analysis we present the predictions in therms of the first two
angular power spectrum coefficients $C_1$ and $C_2$. While the
full--sky constraints shown in Fig.~\ref{exclusion_plot} were imposed using
the sets of coefficients $a_{1m}$ and $a_{2m}$,
the power spectrum $C_l$ reveals the overall sensitivity of certain harmonic to
the respective theoretical anisotropy pattern. In this sense the predictions
shown are conservative. From the Figs.~\ref{proposal_r1}--\ref{proposal_fullsky} one can learn that the large energies area
is most opportune for the DM decay anisotropy search, while at the energies of $\sim 10$~EeV 
the sensitivity need to be increased by at least several hundred times comparing to the recent searches.

\section{Discussion}
\label{discussion}
The obtained results indicate that current EAS experiments are more sensitive
to the photons from DM decay than to the respective anisotropy.
This occurs due to relatively the good
hadron--photon primaries separation in EAS analysis and due to insufficient sensitivity
of the ground based EAS experiments to the large scale anisotropy. A natural obstacle
here is the necessity to combine  the results of two experiments for the full--sky analysis.
We should also note the connection between the anisotropy and gamma--ray signal.
The large--scale anisotropy if observed at a particular energy not accompanied by the gamma--rays
should be attributed to physics other than the DM decay, e.g. to the imprint of
Large Scale Structure of the Universe~\cite{Sigl:2003ay, Sigl:2004yk, Kalashev:2007ph, Koers:2008ba}
or the anisotropic particle acceleration in the local cosmos~\cite{Ahlers:2016njd, Ahlers:2013ima, Fujita:2016yvk}.
In other words, until the gamma--rays of the respective energies are detected the DM
signal should not interfere with the study of astrophysical large--scale anisotropy.

Some of the future experiments may be more sensitive to anisotropy than to gamma--ray DM signal.
For instance the EUSO experiment is planned to have high sensitivity to anisotropy\cite{dOrfeuil:2014qgw} while
its ability of photon--hadron primaries separation is expected to be
lower that in current experiments~\cite{Dawson:2017rsp}. In Ref~\cite{Marzola:2016hyt} the anisotropy
detection prospects from the DM decay signal allowed by current photon limits were found
favourable for EUSO experiment at ultra--high energies. At the same time planned photon sensitivity
improvements in the currently running experiments - Pierre Auger and Telescope Array
(see Ref.\cite{Karg:2015gxa} for details) would make them even more effective in search
for the signal of heavy decaying DM.

\section*{Acknowledgements}
We would like to thank V.~Rubakov, S.~Troitsky and G.~Rubtsov for helpful discussions.
We are especially indebted to R.~Aloisio, V.~Berezinsky and M.~Kachelriess for providing
the numerical code solving the DGLAP equations.
This work has been supported by the Russian Science Foundation grant 14-22-00161.

\suppressfloats

\bibliographystyle{unsrturl}
\bibliography{ref}

\end{document}